\input{aipcheck}

\documentclass[sort&compress]{aipproc}

\layoutstyle{8x11double}

\newcommand{\be}{\begin{equation}}
\newcommand{\ee}{\end{equation}}

\newcommand{\dlt}{\delta}

\newcommand{\br}{{\bf r}}
\newcommand{\bd}{{\bf d}}
\newcommand{\bj}{{\bf j}}
\newcommand{\bE}{{\bf E}}
\newcommand{\bP}{{\bf P}}
\newcommand{\bfe}{{\bf e}}
\newcommand{\bH}{{\bf H}}
\newcommand{\bJ}{{\bf J}}
\newcommand{\bA}{{\bf A}}
\newcommand{\bB}{{\bf B}}
\newcommand{\bS}{{\bf S}}
\newcommand{\bt}{\beta}
\newcommand{\vp}{\varphi}

\newcommand{\al}{\alpha}
\newcommand{\ra}{\rightarrow}

\newcommand{\gm}{\gamma}
\newcommand{\om}{\omega}

\newcommand{\lbd}{\lambda}

\newcommand{\rgl}{\rangle}
\newcommand{\lgl}{\langle}

\begin{document}

\title{Coherent Radiation by Quantum Dots and \\ Magnetic Nanoclusters}


\author{V.I. Yukalov}{
  address={Bogolubov Laboratory of Theoretical Physics, 
Joint Institute for Nuclear Research, Dubna 141980, Russia}
}

\author{E.P. Yukalova}{
  address={Laboratory of Information Technologies, 
Joint Institute for Nuclear Research, Dubna 141980, Russia}
}

\begin{abstract}
The assemblies of either quantum dots or magnetic nanoclusters are 
studied. It is shown that such assemblies can produce coherent radiation. A method 
is developed for solving the systems of nonlinear equations describing the dynamics 
of such assemblies. The method is shown to be general and applicable to systems of 
different physical nature. Despite mathematical similarities of dynamical equations, 
the physics of the processes for quantum dots and magnetic nanoclusters is rather 
different. In a quantum dot assembly, coherence develops due to the Dicke effect of 
dot interactions through the common radiation field. For a system of magnetic clusters, 
coherence in the spin motion appears due to the Purcell effect caused by the feedback 
action of a resonator. Self-organized coherent spin radiation cannot arise without 
a resonator. This principal difference is connected with the different physical nature 
of dipole forces between the objects. Effective dipole interactions between the radiating 
quantum dots, appearing due to photon exchange, collectivize the dot radiation. While 
the dipolar spin interactions exist from the beginning, yet before radiation, and on 
the contrary, they dephase spin motion, thus destroying the coherence of moving spins. 
In addition, quantum dot radiation exhibits turbulent photon filamentation that is 
absent for radiating spins.  
\end{abstract}

\maketitle

\section{Introduction}

The aim of this article is to demonstrate the feasibility of employing nano-objects, 
such as quantum dots and magnetic nanoclusters, as sources of coherent radiation.  
The assemblies of such nano-objects have many features common with the systems of
resonance atoms that are in the basis of laser physics \cite{1,2}, hence, these 
nano-objects also could be used for creating collective sources of coherent radiation, 
similar to resonance atoms. 

Though both, quantum dots and magnetic nanoclusters, remind finite-level atoms and show
several properties common for atoms, exhibiting similarities in the dynamics of dots and 
clusters, there are as well important differences caused by the different physical nature 
of the nano-objects. We concentrate our attention on the collective effects in the 
dynamics of quantum dots and magnetic nanoclusters. We emphasize the necessity of 
employing microscopic description of the processes, where all system parameters are well
defined. The use of phenomenological equations often leads to wrong conclusions that are 
widespread in literature. The theory, based on microscopic Hamiltonians, allows us to 
clearly understand the physical origins of and the necessary conditions for achieving 
coherent dynamics. The microscopic quantum picture is especially important for describing 
the self-organized birth of coherence from initial chaotic fluctuations. Such a 
self-organized coherence cannot be described by phenomenological equations of semi-classical 
type. 

Despite the very different physical nature of quantum dots and magnetic nanoclusters, 
their dynamics can be reduced to mathematically similar equations that can be analyzed 
by the general method of scale separation. At the same time, because of their different 
physical nature, the physics of their collective dynamics is very different. We emphasize 
that the origins of collective phenomena in quantum dots and magnetic nanoclusters are 
{\it principally different}. In a quantum-dot system, coherence develops due to the 
photon exchange through the common radiation field, that is, due to the {\it Dicke effect} 
\cite{3}. While in an assembly of magnetic nanoclusters, the Dicke effect is impossible 
and coherence can arise only through a resonator feedback field, that is, due to the 
{\it Purcell effect} \cite{4}.

\section{Quantum Dots}

Electrons in quantum dots are confined in all three spatial dimensions, which makes their 
spectrum discrete \cite{5,6,7,8}. Exciting an electron from the ground-state level creates 
a hole. The interacting pair of an electron and a hole forms an exciton, whose recombination 
is accompanied by electromagnetic radiation, in a close analogy with atomic radiation. 

There exists a variety of quantum dots, for instance, the dots based on self-assembled
heterostructures, such as InAs/GaAs, InGaAs/GaAs, InGaAs/AlGaAs, GaInAsP/InP, InAs/InP,
InAs/GaInAs, AlInAs/AlGaAs, InP/GaInP, AlGaAs/GaAs, CdSe/ZnSe, and so on. In each dot
there can be between $2$ to $10^5$ electrons. 

The characteristic lengths related to quantum dots are as follows. The dot size is 
$r_{dot} \sim 10^{-7} - 10^{-6}$ cm, the interdot distance, $a \sim 10^{-5} - 10^{-4}$ cm,
radiation wavelength, $\lambda \sim 10^{-4}$ cm, the collection of quantum dots, forming
a kind of a laser, has the radius and length $R, L \sim 10^{-3} - 10^{-2}$ cm, which makes
the volume $V = \pi R^2 L \sim 10^{-8} - 10^{-5}$ cm$^3$. The dot density in a laser is 
$\rho \sim 10^{13} - 10^{17}$ cm$^{-3}$. Hence the total number of dots is
$N = \rho V \sim 10^5 - 10^{12}$. 

The longitudinal relaxation time is due to electron-phonon coupling that is suppressed 
at low temperatures, so that, at helium temperatures, $T_1 \sim 10^{-9}$ s. The 
homogeneous dephasing time is $T_2 \sim 10^{-13} - 10^{-12}$ s. For high-quality 
self-assembled heterostructures, the inhomogeneous broadening is of the same order as
the homogeneous broadening, so that $T^*_2 \sim 10^{-13} - 10^{-12}$ s. 

The transition frequency is $\omega_0 \sim 10^{15}$ Hz. The natural width is 
$\gamma_0 \sim 10^{10}$ Hz, the longitudinal relaxation width, $\gamma_1 \sim 10^9$ Hz,
the homogeneous broadening, $\gamma_2 \sim 10^{12} - 10^{13}$ Hz, and the inhomogeneous
broadening is $\gamma^*_2 \sim 10^{12} - 10^{13}$ Hz. To select and enhance the chosen 
mode, the sample is placed into a high-quality resonator cavity. 

Since the wavelength is much larger than the dot size, $\lambda \gg r_{dot}$, the 
interaction of a dot with electromagnetic field can be treated in the dipole 
approximation. The wavelength is also larger than the interdot distance, $\lambda \gg a$,
which tells us that there should exist essential interaction between dots. And the 
wavelength is much shorter than the sample sizes, $\lambda \ll R, L$, because of which 
the point-sample approximation is not allowed.      
  
The possibility of coherent quantum dot radiation has been mentioned in Ref. \cite{9} 
and discussed in detail in Ref. \cite{10}. The microscopic Hamiltonian for an ensemble 
of radiating quantum dots in a semiconductor matrix inside a resonator cavity \cite{10} 
is the sum
\be
\label{1}
 \hat H = \hat H_d +  \hat H_f + \hat H_{df} + \hat H_{mf} \; .
\ee
Here the dot Hamiltonian is
\be
\label{2}
 \hat H_d = \sum_{i=1}^N \om_0 \left ( \frac{1}{2} + S_i^z \right ) \;  ,
\ee
with $\omega_0$ being the carrying transition frequency and $S^z_i$, a pseudospin 
operator of population imbalance. The field Hamiltonian is
\be
\label{3}
 \hat H_f = \frac{1}{8\pi} \int \left ( \bE^2 + \bH^2\right ) d\br \; ,
\ee
where ${\bf E}$ is electric field, $H = \nabla \times {\bf A}$ is magnetic field,
${\bf A}$ is vector potential satisfying the Coulomb calibration 
$\nabla \cdot {\bf A} = 0$. The dot-field interaction is given by the Hamiltonian
\be
\label{4}
 \hat H_{df} = -\sum_{i=1}^N \left ( \frac{1}{c} \; \bJ_i \cdot \bA_i +\bP_i
\cdot \bE_{i0} \right ) \;  ,
\ee
with the transition current
\be
\label{5}
 \bJ_i = i \om_0 \left (\bd S_i^+ - \bd^* S_i^- \right ) \; ,  
\ee
and transition polarization
\be
\label{6}
 \bP_i =  \bd S_i^+ + \bd^* S_i^- \; ,
\ee
where ${\bf A}_i = {\bf A}({\bf r}_i, t)$, ${\bf E}_{i0}$ is the cavity seed field, 
${\bf d}$, transition dipole, and $S^{\pm}_i$ is a pseudospin ladder operator. The 
cavity is filled by a semiconducting material interacting with the radiation field 
through the Hamiltonian
\be
\label{7}
 \hat H_{mf} = -\; \frac{1}{c} \; \int \bj_{mat} \cdot \bA \; d\br \;  ,
\ee
where ${\bf j}_{mat}$ is a fluctuating local-density current of the semiconductor 
matrix.     
 
We eliminate the field variables by writing down the Heisenberg equations of motion, 
solving the d'Alembert equation for ${\bf A}_i$, expressing it through the pseudospin 
operators, and substituting the d'Alembert solution to the pseudospin equations of 
motion \cite{10,11}. Then we average the latter equations defining the statistical 
average
$$
 \lgl S_i^\al(t) \rgl \equiv {\rm Tr}\hat\rho(0) S_i^\al(t) \;  .
$$
The resulting equations are written for the {\it transition function}
\be
\label{8}
  u_i \equiv 2 \lgl S_i^- \rgl \; ,
\ee
{\it coherence intensity}
\be
\label{9}
 w_i \equiv \frac{2}{N} \sum_{j(\neq i)} \lgl S_i^+ S_j^- +
S_j^+ S_i^- \rgl \;  ,
\ee
and the {\it population imbalance}
\be
\label{10}
 s_i \equiv 2 \lgl S_i^z \rgl \;  .
\ee

\section{Scale Separation}

The evolution equations for the above variables acquire the form
$$
\frac{du_i}{dt} = f_{iu} \; , \qquad \frac{dw_i}{dt} = f_{iw} \; , \qquad
\frac{ds_i}{dt} = f_{is} \;   ,
$$
with the right-hand sides depending on the variables $u_i, w_i, s_i, \xi_i$, 
and time $t$. Here $\xi_i$ is a random field caused by vacuum fluctuations, dipole
fluctuations, and semiconductor-matrix current fluctuations. We notice that when
the widths $\gamma_0, \gamma_1, \gamma_2$, and $\gamma^*_2$, all of which are smaller 
than $\omega_0$, tend to zero, then $f_{iw}$ and $f_{is}$ also tend to zero, 
while $f_{iu}$ remains finite. This allows us to classify the variable $u_i$ as 
fast and the variables $w_i, s_i$ as slow. Therefore, for solving these evolution 
equations, we can resort to the scale separation approach \cite{11,12,13,14,15}. 
The solution proceeds as follows. We solve the equation for the fast variable $u_i$, 
keeping the slow variables fixed. This solution is substituted into the right-hand 
sides of the equations for the slow variables. Then these right-hand sides are 
averaged over time and over the random fluctuations $\xi_i$, again keeping the 
slow variables fixed, which corresponds to the rule
$$
 \overline f \equiv 
\lim_{\tau\ra\infty} \; \frac{1}{\tau} \int_0^\tau \lgl\lgl f(t) \rgl\rgl dt \;  ,
$$
where the double angle brackets denote the averaging over the stochastic fluctuations.
Thus we come to the equations
$$
 \frac{dw_i}{dt} = \overline f_{iw} \; , \qquad 
\frac{ds_i}{dt} = \overline f_{is} \; ,  
$$
defining the guiding centers of the slow variables.

\section{Dot Radiation}

\vskip 3mm
The total radiation intensity of quantum dots, averaged over fast fluctuations, 
can be represented \cite{16} as the sum of two terms,
\be
\label{11}
 I(t) = I_{inc}(t) + I_{coh}(t) \;  ,
\ee
where the first term is the incoherent radiation intensity
\be
\label{12}
I_{inc}(t) = \frac{1}{2}\; \om_0 \gm_0 ( 1+ s) N \; ,
\ee
while the second term is the coherent radiation intensity
\be
\label{13}
 I_{coh}(t) = \om_0 \gm_0 \vp_s w N^2 \; ,
\ee
with $\varphi_s$ being a shape factor \cite{1} and $s$ and $w$ being the spatial
averages over the coherence volume \cite{10}.        

The spatial distribution of radiation depends on the value of the Fresnel
number $F \equiv R^2/\lbd L$. For $F < 1$, the radiating beam is uniform. In 
the range $1 < F < 10$, the radiation is separated into a few Gauss-Laguerre 
modes dictated by the sample geometry. And when $F>10$, the effect of turbulent 
photon filamentation \cite{17,18,19,20,21} appears. Then the radiation beam separates 
into the large number 
\be
\label{14}
N_f = 3.3 F 
\ee
of filaments with radius
\be
\label{15}
r_f = 0.3 \sqrt{\lbd L} \;   .
\ee
 
According to Eq. (13), the coherent radiation is formally proportional to the number
of radiators squared, $N^2$. However, since $\lambda \ll R, L$, it is necessary to 
take into account the shape factor that, depending on either the pencil or disk 
geometry of the sample, behaves as
\begin{eqnarray}
\nonumber
\vp_s \propto \left \{ \begin{array}{ll}
N^{-1/3} & ~~~ (pencil) \\
N^{-2/3} & ~~~ (disk)
\end{array} \right. \; .
\end{eqnarray}
Respectively, the coherent radiation exhibits the dependence on the dot number as 
\begin{eqnarray}
\nonumber
I_{coh} \propto \left \{ \begin{array}{ll}
N^{5/3} & ~~~ (pencil) \\
N^{4/3} & ~~~ (disk)
\end{array} \right. \; .
\end{eqnarray}

The temporal evolution of radiation consists of the following stages. The very 
first and short is the {\it interaction stage}, when the dots start radiating, but 
have had yet no time for initiating mutual interactions:
\be    
\label{16}
 0 < t < t_{int} \qquad (interaction \; stage) \;  .
\ee
Here the {\it interaction time} is 
$$
 t_{int} = \frac{a}{c} \sim 10^{-15} - 10^{-14}\; {\rm s} \;  .
$$

The second is the {\it quantum stage}, when the dots have started interacting with 
each other through the common radiation field, but radiate yet chaotically, without
noticeable mutual correlations:
\be
\label{17}
 t_{int} < t < t_{coh} \qquad (quantum \; stage) \;  .
\ee
The stage lasts till the {\it coherence time}
$$
t_{coh} = \frac{T_2}{2gs_0} \sim 10^{-14} - 10^{-13} {\rm s} \;   ,
$$
in which $s_0 \equiv s(0)$ and the {\it coupling parameter}
$$ 
g =\frac{\rho\gm_0\lbd^2 L}{4\pi\gm_2}
$$ 
characterizes the effective dot interactions 
through the photon exchange. With the considered parameters, the value of $g$ 
can be as high as $g \sim 10^3$ for the dot density $\rho \sim 10^{17}$ cm$^{-3}$. 
The interaction and quantum stages cannot be described by the semiclassical 
approximation, but require microscopic quantum description. 

After the coherence time, the dots become correlated and radiate coherently in 
the temporal interval
\be
\label{18}
 t_{coh} < t < T_2 \qquad (coherent\; stage) \;  ,
\ee
till they are dephased at the dephasing time $T_2 \sim 10^{-13} - 10^{-12}$ s.
In the case of pure superradiance, when the coherent radiation is self-organized,
the radiation peak occurs at the {\it delay time} $t_0 \sim 5 t_{coh}$ and the 
superradiance pulse duration is $t_p \sim 2 t_{coh}$. 

The next is the {\it relaxation stage}, when the radiation can yet be noticeable, 
but already not coherent and quickly diminishing:
\be
\label{19}
  T_2 < t < T_1 \qquad (relaxation \; stage) \; .
\ee
The radiation practically vanishes at the longitudinal relaxation time 
$T_1 \sim 10^{-9}$ s. 

Finally, after $T_1$, there is almost no radiation, but just some weak splashes 
caused by random fluctuations. This can be called the {\it stationary stage}:
\be
\label{20}
 t > T_1 \qquad (stationary \; stage) \;  .
\ee

Figure 1 presents the typical behavior of the coherence intensity and population 
imbalance as functions of time. These functions are smoothed by averaging over fast 
oscillations of the period $2 \pi / \omega_0$.  

\begin{figure}[ht]
\centerline{\includegraphics[width=8.5cm]{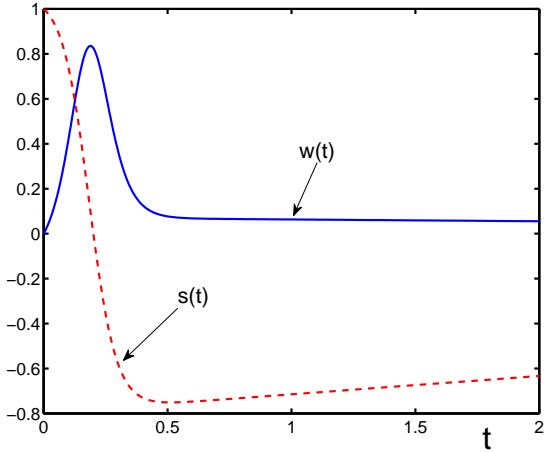} }
\caption{Pure dot superradiance, with initial conditions $w_0=0$, $s_0=1$. 
Coherence intensity (solid line) and population difference (dashed line) as 
functions of time (in units of $T_2$) for  $\gm_1=0.003\gm_2$ and  $g=10$.}
\label{fig:Fig.1}
\end{figure}

If the dots are subject to stationary nonresonant pumping, characterized by the 
pumping parameter $\gamma_1^*$, then the regime of {\it pulsing superradiance} 
develops, exhibiting several superradiant pulses, as is shown in Fig. 2.

\begin{figure}[ht]
\centerline{\includegraphics[width=8.5cm]{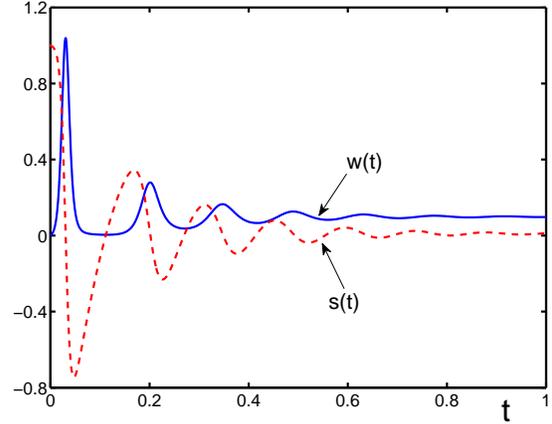} }
\caption{Pulsing dot superradiance, under initial conditions $w_0=0$, $s_0=1$, 
with the pumping parameter $\gm^*_1 =10\gm_2$ and coupling parameter $g=100$. 
Coherent intensity (solid line) and population difference (dashed line). }
\label{fig:Fig.2}
\end{figure} 

Figures 1 and 2 correspond to pure dot superradiance, when there are no external 
pulses imposing initial coherence on the sample, so that $w_0 \equiv w(0) = 0$. 
The radiation can also be pushed by a coherent initial pulse yielding $w_0 > 0$.
Then one has the triggered dot superradiance, with essentially shortened delay 
time $t_0$. 

To estimate the radiation intensity in dimensional units (in Watts), we take the 
typical dot parameters mentioned above, which gives the incoherent radiation 
intensity $I_{inc} \sim 10^{-9} N$ W and the coherent radiation intensity 
$I_{coh} \sim 10^{-9} \varphi_s N^2$ W. The shape factor for the pencil and disk
geometry is
\begin{eqnarray}
\nonumber
\vp_{s} \simeq \left \{ \begin{array}{ll}
\frac{3\lbd}{8L} \; , & ~~~ \frac{R}{L} \ll 1  \\
\\
\frac{3}{8} \left ( \frac{\lbd}{\pi R}\right )^2 \; , & ~~~ \frac{L}{R} \ll 1 \; ,
\end{array} \right.
\end{eqnarray}
which, with the given parameters, translates into 
\begin{eqnarray}
\nonumber
\vp_{s} \simeq \left \{ \begin{array}{ll}
10^{-2} \; , & ~~~ \frac{R}{L} \ll 1  \\
\\
10^{-3}\; , & ~~~ \frac{L}{R} \ll 1 \; .
\end{array} \right.
\end{eqnarray}
The number of dots that can radiate coherently is 
$$
N_{coh}=\rho \pi \lbd^2 L \; .
$$
For the dot density $\rho \sim 10^{17}$ cm$^{-3}$, we have $N \sim 10^7$. 
Then, for a pencil-like sample of $N \sim 10^7$ dots, we get the incoherent 
intensity $I_{inc} \sim 10^{-2}$ W and the coherent intensity 
$I_{coh} \sim 10^3$ W. The superradiant pulse duration is $10^{-13}$ s.

\section{Magnetic Nanoclusters}

Under magnetic nanoclusters, we here keep in mind three possible types of such
objects, polarized nanomolecules, magnetic nanomolecules, and magnetic nanoclusters
as such. Polarized nanomolecules can be put together, forming sufficiently large
solids. Magnetic nanomolecules can form crystals with well defined crystalline 
lattice. Magnetic nanoclusters are the clusters that can possess a total nonzero 
spin, thus enjoying nonzero magnetization. The sizes of such clusters are of 
nanoscale, having typical radii not larger than the coherence radius, since only then 
a cluster presents a single magnetic domain with a nonzero total spin. The larger 
clusters separate into several domains, so that the total spin becomes zero. 
There exists a variety of different magnetic nanoclusters \cite{22,23,24,25,26}. 
It is possible to distinguish three main classes of such magnetic objects. 

(i) {\it Polarized nanomolecules}. These are large molecules that do not possess 
nonzero spin in their ground state. However, they contain many hydrogen atoms whose 
protons can be polarized and, being kept a low temperature $T < 1$ K, the polarization 
remains frozen for very long time. There exist many such molecules, among which we 
can mention propanediol C$_3$H$_8$O$_2$, butanol C$_4$H$_9$OH, and ammonia NH$_3$. 
Materials, formed by such molecules, are characterized by the density of protons, 
each having spin $S = 1/2$. 

(ii) {\it Magnetic nanomolecules}. Such molecules possess nonzero total spin, due to
electrons, in their ground state. At low temperature, below the blocking temperature
$T_B \sim 1 - 10$ K, the spin can be fixed for long time. There is a number of such
magnetic molecules. The most often considered are the molecules denoted as Mn$_{12}$ 
and Fe$_8$, whose complete formulas are
$$
{\rm
Mn_{12}O_{12}(CH_3COO)_{16}(H_2O)_4(2CH_3COOH) 4H_2 0 } \;   , 
$$
$$
{\rm
[Fe_8O_2(OH)_{12} tacn_6 ]^{+8} } \; ,
$$
where {\it tacn} stands for triazacyclononane. The ground-state spin of these molecules 
is $S = 10$. Magnetic nanomolecules can form crystals with a good crystalline lattice.       

(iii) {\it Magnetic nanoclusters}. Truly magnetic nanoclusters are composed of magnetic
atoms or molecules correlated by exchange interactions. For instance, such clusters 
can be formed by the atoms of Fe, Ni, or Co, or by oxides, such as NiO, Fe$_2$O$_3$,
and NiFe$_2$O$_4$. Below the blocking temperature $T_B \sim 10 - 100$ K, the total spin
of each nanocluster, reaching $S \sim 10^2 - 10^5$, can be kept frozen for very long 
time.      

Magnetic nanoclusters find numerous applications in magnetic chemistry, biomedical 
imaging, cancer treatment, genetic engineering, waste cleaning, information storage,
and quantum computing \cite{22,23,24,25,26}.

The generic Hamiltonian for a system of magnetic nanoclusters writes as
\be
\label{21}
 \hat H = \sum_i \hat H_i + \frac{1}{2} \sum_{i\neq j} \hat H_{ij} \; ,
\ee 
where the index $i = 1,2,\ldots,N$ enumerates nanoclusters. The single-cluster
Hamiltonian is
$$
\hat H_i = -\mu_i \bB \cdot \bS_i - D(S_i^z)^2 + D_2 (S_i^x)^2 +
$$
\be
\label{22}
   +
D_4 \left [ (S_i^x)^2  (S_i^y)^2  +  (S_i^y)^2  (S_i^z)^2 +
(S_i^z)^2  (S_i^x)^2 \right ] \; ,
\ee
where the first term is the Zeeman energy and the following terms describe magnetic 
anisotropy. The nanocluster interactions are given by the Hamiltonian
\be
\label{23}
 \hat H_{ij} = \sum_{\al\bt} D_{ij}^{\al\bt} S_i^\al S_j^\bt \;  ,
\ee
with the dipolar tensor
\be
\label{24}
D_{ij} = \frac{\mu_i\mu_j}{r_{ij}^3} \; \left ( \dlt_{\al\bt} -
 3 n_{ij}^\al n_{ij}^\bt \right )   ,
\ee
in which $r_{ij}\equiv |\br_{ij}|$, ${\bf n}_{ij}\equiv \br_{ij}/r_{ij}$, and
$\br_{ij} \equiv \br_i - \br_j$.

The total magnetic field 
\be
\label{25}
\bB = B_0 \bfe_z + H \bfe_x
\ee
consists of an external magnetic field $B_0$ and of the feedback field of a 
resonant electric circuit \cite{25,26}. The resonator feedback field is described
by the Kirchhoff equation
\be
\label{26}
\frac{dH}{dt} + 2\gm H + \om^2 \int_0^t H(t') \; dt' = 
- 4\pi\eta \; \frac{dm_x}{dt} \;   ,
\ee
where $\gamma$ is resonator damping, $\omega$, resonator natural frequency, 
$\eta$ is a filling factor and 
$$
m_x \equiv \frac{1}{V} \sum_j \mu_j \lgl S_j^x \rgl \; .   
$$

Similarly to Eqs. (8) to (10), we introduce the dynamical variables, the 
{\it transverse component}
\be
\label{27}
 u \equiv \frac{1}{SN} \sum_{j=1}^N \lgl S_j^- \rgl \;  ,
\ee
the {\it coherence intensity}
\be
\label{28}
 w \equiv \frac{1}{S^2N^2} \sum_{i\neq j}^N \lgl S_i^+ S_j^- \rgl \;   ,
\ee
and the {\it spin polarization}
\be
\label{29}
s \equiv \frac{1}{SN} \sum_{j=1}^N \lgl S_j^z \rgl \;   .
\ee

The evolution equations for these variables are obtained from the related 
Heisenberg equations for spins, with the feedback field given by the Kirhhoff
equation (26). The following steps of solving these equations are analogous to 
those corresponding to the solution of pseudospin equations for quantum dots. 
These equations have been analyzed for polarized molecules 
\cite{11,12,13,14,15,25,27}, magnetic nanomolecules \cite{25,28,29,30,31,32}, 
and for magnetic nanoclusters \cite{26,33,34}. The results obtained by the scale 
separation approach \cite{11,12,13,14,15,25} have been compared with direct 
numerical simulations \cite{35,36,37} of the evolution equations for spins, 
both ways being in good agreement with each other. 

Here we illustrate the solution to the spin equations of motion for the 
parameters typical of such nanoclusters as those composed of Fe, Ni, and Co.
These nanoclusters are characterized by the following parameters. The Zeeman
frequency is taken as $\omega_0 \sim 10^{10} - 10^{12}$ Hz, which translates 
into the wavelength $\lambda \sim 0.1 - 10$ cm. The sample of 
$N \sim 10^{14} - 10^{20}$ nanoclusters has the linear sizes comparable with 
the wavelength, because of which there is no the filamentation effect that
occurs for quantum dots. The nanocluster density in the sample is 
$\rho \sim 10^{20}$ cm$^{-3}$. The intercluster distance is $a \sim 10^{-7}$ cm. 
The feedback rate, due to the coupling of the nanoclusters with a resonator, 
is $\gamma_0 \sim 10^{10}$ s$^{-1}$. Typical anisotropy parameters are given
by the relations
$$  
\frac{D}{\hbar \gm_0} \sim 10^{-3} \; , \qquad   
\frac{D_2}{\hbar \gm_0} \sim 10^{-3} \; , \qquad 
\frac{D_4}{\hbar \gm_0} \sim 10^{-10} \;  .
$$
Below the blocking temperature $T_B \sim 10$ K, the total spin $S \sim 10^3$ is 
frozen, so that the longitudinal relaxation time $T_1$ reaches years. The 
transverse dephasing time $T_2 \sim 10^{-10}$ s is caused by dipole spin 
interactions. The coupling parameter, defining the effective spin-resonator 
interaction, is $g \sim \omega_0/ \gamma \sim 1 - 100$.       

Figure 3 demonstrates the temporal behaviour of the coherence intensity and spin
polarization in the regime of pure spin superradiance.

\begin{figure} [ht]
\centering
\begin{tabular}{lr}
\includegraphics[width=8.5cm]{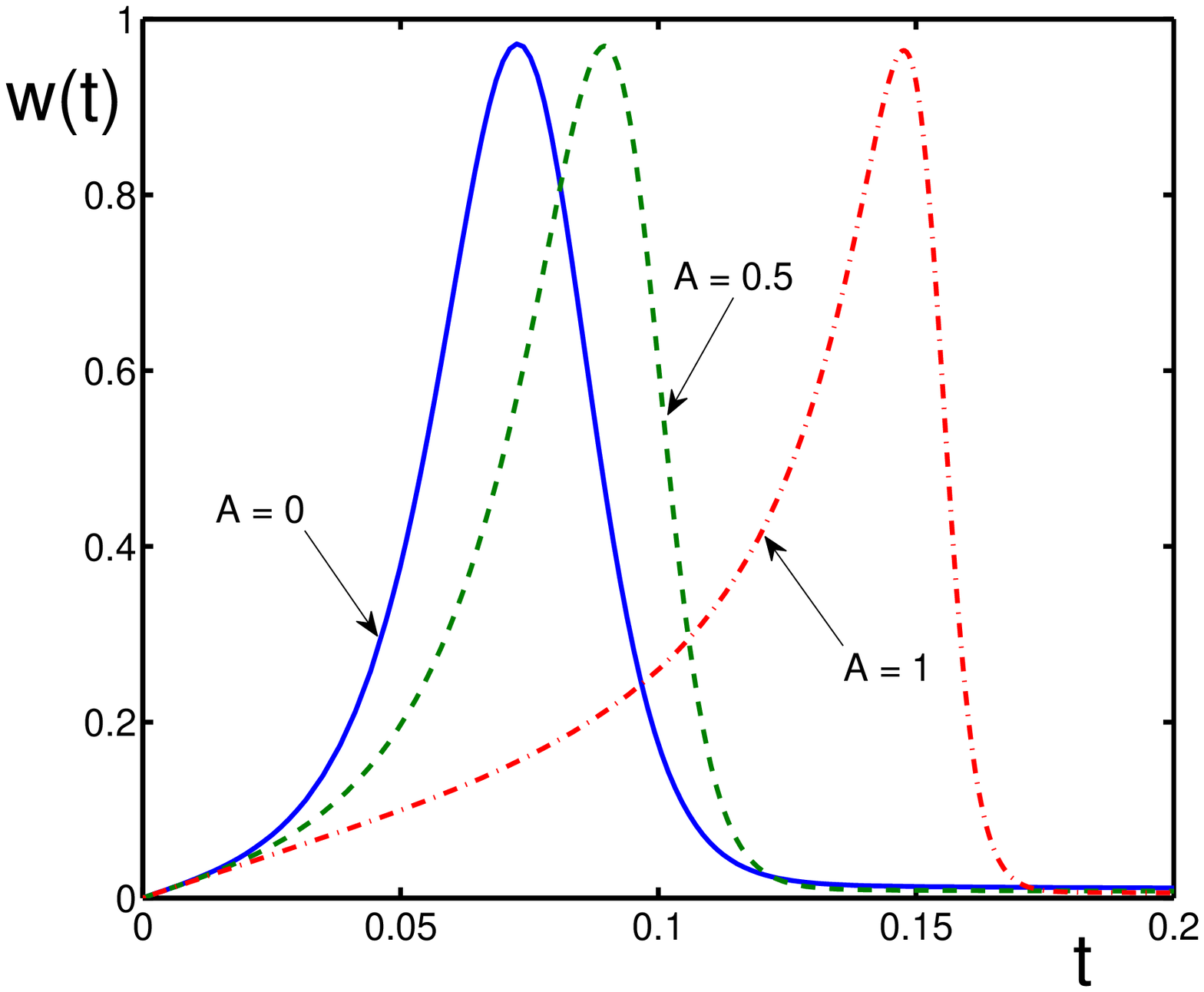}  &
\includegraphics[width=8.5cm]{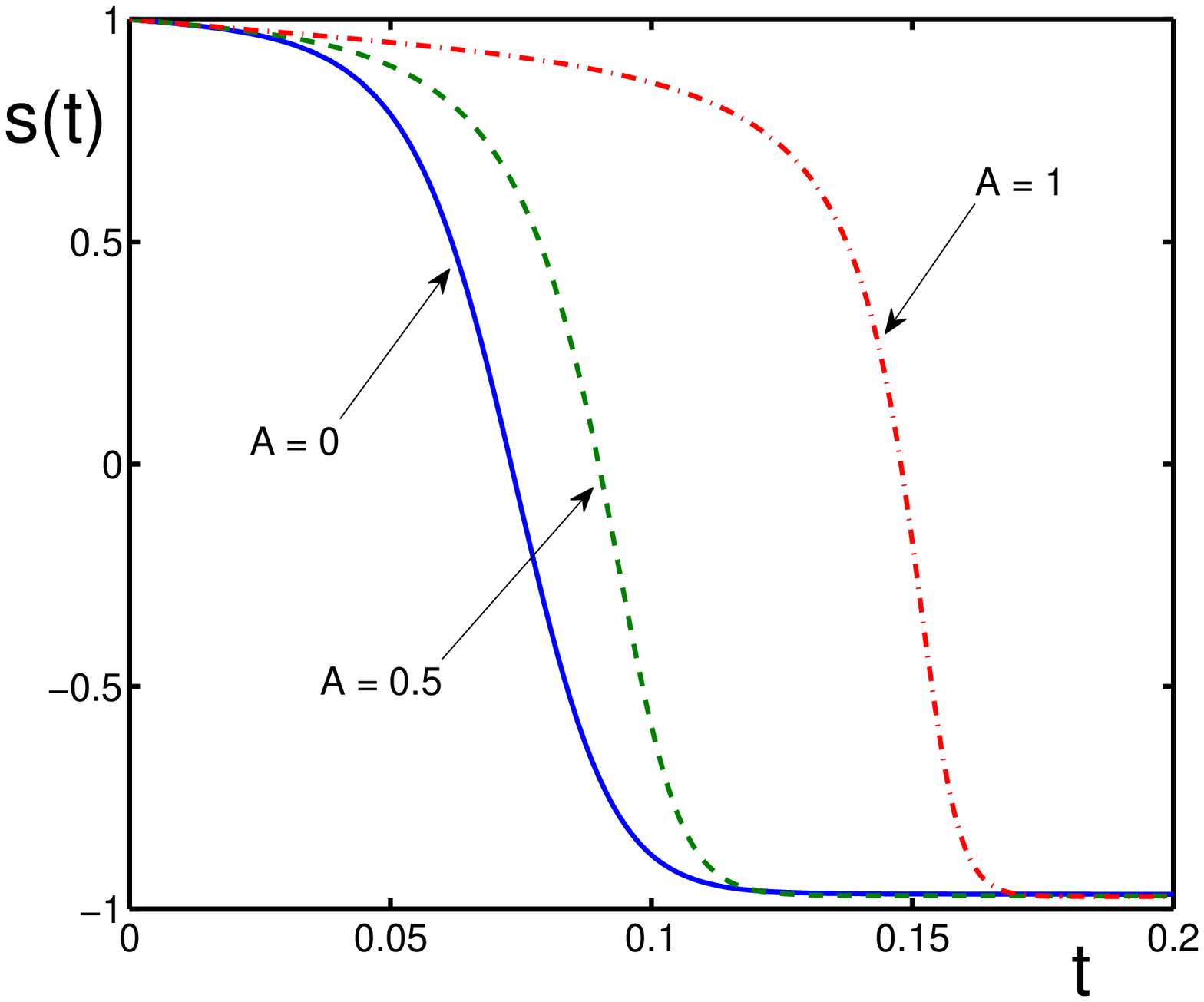} 
\end{tabular}
\caption{Pure spin superradiance, under initial conditions $w_0=0$, $s_0=1$. 
Coherence intensity $w(t)$ and spin polarization $s(t)$as functions of time 
(in units of $1/\gm_2$) for $\gm=10\gm_2$, $\gm_1=10^{-3}\gm_2$, $g = 100$, 
and for different anisotropy parameters: $A = 0$ (solid line), $A=0.5$ 
(dashed line), and $A=1$ (dashed-dotted line).
}
\label{fig:Fig.3}
\end{figure}

\section{Nanocluster Radiation}

The total intensity of spin radiation for the system of magnetic nanoclusters
is the dipole radiation intensity
$$
 I_S(t) = 
\frac{2\mu_0^2}{3c^3} \left | \sum_i \lgl \ddot{\bS}(t) \rgl \right |^2 \;  ,
$$
where, as usual, the dots mean time differentiation. This yields
\be
\label{30}
I_S(t) = \frac{2\mu_0^2\om_0^4}{3c^3} \; N^2 S^2\; [ 1 - As(t) ]^4 w(t) \;  .
\ee
Here $\mu_0$ is the average nanocluster magnetic moment and $A$ is the anisotropy 
parameter defined by the expressions  
$$
A \equiv \frac{\om_A}{\om_0} \; , \qquad
\om_A \equiv \om_D + \frac{1}{2} \; \om_2 \; , 
$$
$$ 
\om_D \equiv (2S-1)D \; , \qquad \om_2 \equiv (2S -1 )D_2 \;  .
$$

The superradiant pulse is accompanied by the spin reversal. The reversal 
time is $t_{rev} \sim 10^{-12}$ s. The peak of radiation intensity, depending 
on the number of nanoclusters $N \sim 10^{14} - 10^{20}$, reaches
$I_S \sim 1 - 10^{12}$ W. The typical superradiant pulse is shown in Fig. 4, 
where $I(t)$ is presented in dimensionless units, so that 
$$
I(t)\equiv [1-As(t)]^4 w(t) \; .
$$ 

\begin{figure}[ht]
\centerline{\includegraphics[width=8.5cm]{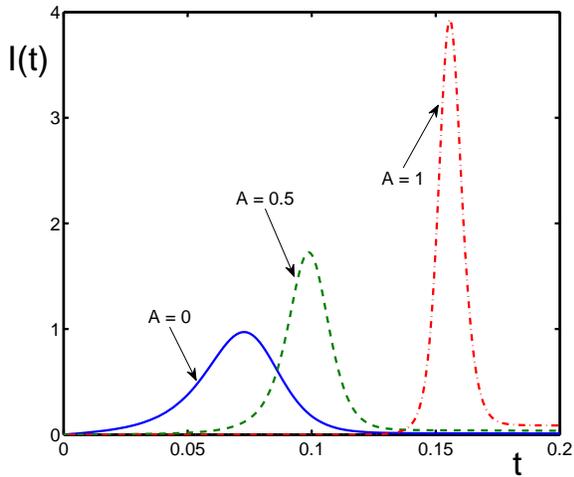} }
\caption{Pure spin superradiance, under initial conditions $w_0=0$, $s_0=1$. 
Dimensionless radiation intensity $I(t)$, as functions of time (in units of 
$1/\gm_2$) for the same parameters as in Fig.3, and for different anisotropy 
parameters: $A = 0$ (solid line), $A=0.5$ (dashed line), and $A=1$ (dashed-dotted 
line).
}
\label{fig:Fig.4}
\end{figure} 

The possibility of obtaining relatively high 
radiation intensity is due to the high spin values of the nanoclusters, 
$S=10^3$.

\section{Conclusion}

We have shown that coherent radiation can be realized with the ensemble of quantum 
dots as well as with the assembly of magnetic nanoclusters. In both the cases, it
is possible to represent the evolution equations in a similar mathematical form,
as the equations of pseudospin or spin variables. Therefore, these equations can be
solved by employing the method of scale separation. By direct numerical simulations 
of the equations of motion it is shown that this approach provides accurate 
description of the dynamic phenomena. 

Despite the formal similarity of the evolution equations, the underlying physics
for quantum dots and magnetic nanoclusters is drastically different. For quantum 
dots, the coherence develops due to the photon exchange through the common radiation
field, that is, due to the {\it Dicke effect}. While for magnetic nanoclusters the arising
coherence is caused by the resonator feedback field, that is, due to the {\it Purcell 
effect}. The coherent spin motion is unachievable for spin systems without a resonator
\cite{28,29,30,31,32,33,34}.

The appearance of coherence in quantum dots has been detected experimentally \cite{38},
though the realization of pure superradiance, to our knowledge, has not been 
accomplished. 
  
There have been attempts to observe radiation from magnetic molecules Mn$_{12}$ 
\cite{39} and Fe$_8$ \cite{40}. However, in these attempts, no resonator was involved. 
But, as is emphasized above, without a resonator no coherence in the spin motion can 
exist. Because of this, such a radiation from neither magnetic molecules nor magnetic 
nanoclusters has yet been experimentally detected.

\begin{theacknowledgments}
Financial support from the Russian Foundation for Basic 
Research is acknowledged. 
\end{theacknowledgments}

\bibliographystyle{aipproc}   

\begin{thebibliography}{9}

\bibitem{1}
L. Allen, and J.~H. Eberly,
 \emph{Optical Resonance and Two-Level Atoms}, Wiley,
  New York, 1975.

\bibitem{2}
A.~V. Andreev, V.~I. Emelyanov, and Y.~A Ilinsky,
 \emph{Cooperative Effects in Optics}, Institute of Physics, 
  Bristol, 1993. 

\bibitem{3}
 R.~H. Dicke, 
\emph{Phys. Rev.} {\bf 93}, 99 (1954). 


\bibitem{4} 
E.~M. Purcell,
\emph{Phys. Rev.} {\bf 69}, 681 (1946).

\bibitem{5}
L.~P. Kouwenhoven, D.~G. Austing, and S. Tarucha,   
\emph{Rep. Prog. Phys.} {\bf 64}, 701 (2001).

\bibitem{6}
S.~M. Riemann and M. Mannien,
\emph{Rev. Mod. Phys.} {\bf 74}, 1283 (2002). 

\bibitem{7}
C. Yannouleas and U. Landman,
\emph{Rep. Prog. Phys.} {\bf 70}, 2067 (2007).

\bibitem{8}
J.~L. Birman, R.~G. Nazmitdinov, and V.~I. Yukalov,
\emph{Phys. Rep.} {\bf 526}, 1 (2013). 
 
\bibitem{9} 
M. Singh, V.~I. Yukalov, and W. Lau,
in \emph{Nanostructures: Physics and Technology}, edited by 
Z. Alferov and L. Esaki,
Ioffe Institute, St. Petersburg, 1998, p. 327. 

\bibitem{10} 
V.~I. Yukalov and E.P. Yukalova,
\emph{Phys. Rev. B} {\bf 81}, 075308 (2010). 

\bibitem{11}
V.~I. Yukalov and E.~P. Yukalova,
\emph{Phys. Part. Nucl.} {\bf 31}, 561 (2000). 

\bibitem{12} 
V.~I. Yukalov,
\emph{Laser Phys.} {\bf 3}, 870 (1993).

\bibitem{13} 
V.~I. Yukalov,
\emph{Phys. Rev. Lett.} {\bf 75}, 300 (1995).

\bibitem{14} 
V.~I. Yukalov,
\emph{Laser Phys.} {\bf 5}, 970 (1995).

\bibitem{15} 
V.~I. Yukalov,
\emph{Phys. Rev. B} {\bf 53}, 9232 (1996).

\bibitem{16} 
V.~I. Yukalov,
\emph{Eur. Phys. J. D} {\bf 13}, 83 (2001). 

\bibitem{17}
 V.~I Yukalov
\emph{J. Mod. Opt.} {\bf 35}, 35 (1988).

\bibitem{18} 
V.~I Yukalov
\emph{J. Mod. Opt.} {\bf 37}, 1361 (1990).

\bibitem{19}
 V.~I. Yukalov,
\emph{Phys. Lett. A} {\bf 278}, 30 (2000).

\bibitem{20} 
V.I. Yukalov,
Phys. Lett. A {\bf 284}, 91 (2001).


\bibitem{21} 
V.~I. Yukalov,
\emph{Physica A} {\bf 291}, 255 (2001). 

\bibitem{22}
B. Barbara, L. Thomas, F. Lionti, I. Chiorescu, and A. Sulpice,
\emph{J. Magn. Magn. Mater.} {\bf 200}, 167 (1999).

\bibitem{23} 
W. Wernsdorfer,
\emph{Adv. Chem. Phys.} {\bf 118}, 99 (2001).

\bibitem{24} 
J. Ferre,
\emph{Topics Appl. Phys.} {\bf 83}, 127 (2002).

\bibitem{25} 
V.~I. Yukalov and E.~P. Yukalova,
\emph{Phys. Part. Nucl.} {\bf 35}, 348 (2004).

\bibitem{26} 
V.~I. Yukalov and E.~P. Yukalova,
\emph{J. Phys. Conf. Ser.} {\bf 393}, 012004 (2012).

\bibitem{27}
V.~I. Yukalov,
\emph{Laser Phys.} {\bf 2}, 559 (1992).

\bibitem{28}
 V.~I. Yukalov,
Laser Phys. {\bf 8}, 1089 (2002).

\bibitem{29}
V.~I. Yukalov,
\emph{Phys. Rev. B} {\bf 71}, 184432 (2005).

\bibitem{30} 
V.~I. Yukalov, and E.~P. Yukalova,
\emph{Eur. Phys. Lett.} {\bf 70}, 306 (2005). 

\bibitem{31} 
V.~I. Yukalov and E.~P. Yukalova,
\emph{Laser Phys. Lett.} {\bf 2}, 302 (2005).

\bibitem{32} 
 V.~I. Yukalov and E.~P. Yukalova,
\emph{Laser Phys. Lett.} {\bf 2}, 356 (2005).

\bibitem{33} 
V.~I. Yukalov and E.~P. Yukalova,
\emph{Laser Phys. Lett.} {\bf 8}, 804 (2011).

\bibitem{34} 
V.~I. Yukalov and E.~P. Yukalova,
\emph{J. Appl. Phys.} {\bf 111}, 023911 (2012). 

\bibitem{35}
T.~S. Belozerova, V.~K. Henner, and V.~I. Yukalov,
\emph{Phys. Rev. B} {\bf 46}, 682 (1992).


\bibitem{36} 
V.~I. Yukalov, V.~K. Henner, P.~V. Kharebov, and E.~P. Yukalova,
\emph{Laser Phys. Lett.} {\bf 5}, 887 (2008).

\bibitem{37} 
P.~V. Kharebov, V.~K. Henner, and V.~I. Yukalov,
\emph{J. Appl. Phys.} {\bf 113}, 043902 (2013). 

\bibitem{38}
M. Scheibner, T. Schmidt, L. Worshech, A. Forchel, G. Bacher, 
T. Passow, and D. Hommel, 
\emph{Nature Phys.} {\bf 3}, 106 (2007). 

\bibitem{39}
M. Bal, J.~R. Friedman, K. Mertes, W. Chen, E.~M. Rumberger, 
D.~N. Hendrickson, N. Avraham, Y. Myasoedov, H. Shtrikman, and 
E. Zeldov,
\emph{Phys. Rev. B} {\bf 70}, 140403 (2004). 

\bibitem{40}
 O. Shafir and A. Keren,
\emph{Phys. Rev. B} {\bf 79}, 180404 (2009). 

\end{thebibliography}

\end{document}